# Operating Regimes of Signaling Cycles: Statics, Dynamics, and Noise Filtering

Carlos Gomez-Uribe,[1,2] George C. Verghese,[2] and Leonid A. Mirny[1]*

[1]Harvard-MIT Division of Health Sciences and Technology,
[2]Department of Electrical Engineering and Computer Science,
Massachusetts Institute of Technology,
77 Massachusetts Avenue, Cambridge, MA 02139 USA
*Corresponding Author

## Abstract

A ubiquitous building block of signaling pathways is a cycle of covalent modification (e.g., phosphorylation and dephosphorylation in MAPK cascades). Our paper explores the kind of information processing and filtering that can be accomplished by this simple biochemical circuit.

Signaling cycles are particularly known for exhibiting a highly sigmoidal (ultrasensitive) input-output characteristic in a certain steady-state regime. Here we systematically study the cycle's steady-state behavior and its response to time-varying stimuli. We demonstrate that the cycle can actually operate in four different regimes, each with its specific input-output characteristics. These results are obtained using the total quasi-steady-state approximation, which is more generally valid than the typically used Michaelis-Menten approximation for enzymatic reactions. We invoke experimental data that suggests the possibility of signaling cycles operating in one of the new regimes.

We then consider the cycle's dynamic behavior, which has so far been relatively neglected. We demonstrate that the intrinsic architecture of the cycles makes them act - in all four regimes - as tunable low-pass filters, filtering out high-frequency fluctuations or noise in signals and environmental cues. Moreover, the cutoff frequency can be adjusted by the cell. Numerical simulations show that our analytical results hold well even for noise of large amplitude. We suggest that noise filtering and tunability make signaling cycles versatile components of more elaborate cell signaling pathways.

## Authors Summary

 A cell is subjected to constantly changing environments and time-varying stimuli. Signals sensed at the cell surface are transmitted inside the cell by signaling pathways. Such pathways can transform signals in diverse ways and perform some preliminary information processing.

A ubiquitous building block of signaling pathways is a simple biochemical cycle involving covalent modification of an enzyme-substrate pair. Our paper is devoted to fully characterizing the static and dynamic behavior of this simple cycle, an essential first step in understanding the behavior of interconnections of such cycles.

It is known that a signaling cycle can function as a static switch, with the steady-state output being an «ultrasensitive» function of the input, i.e., changing from a low to high value for only a small change in the input. We show that there are in fact precisely four major regimes of static and dynamic operation (with the «ultrasensitive» being one of the static regimes). Each regime has its own input-output characteristics.

Despite the distinctive features of these four regimes, they all respond to time-varying stimuli by filtering out high-frequency fluctuations or noise in their inputs, while passing through the lower-frequency information-bearing variations. A cell can select the regime and tune the noise-filtering characteristics of the individual cycles in a specific signaling pathway. This tunability makes signaling cycles versatile components of elaborate cell-signaling pathways.

# Introduction

Cells rely on chemical interactions to sense, transmit, and process time-varying signals originating in their environment. Because of the inherent stochasticity of chemical reactions, the signals transmitted in cell signaling pathways are buried in noise. How can cells then differentiate true signals from noise? We examine this in the context of a basic but ubiquitous module in signaling cascades: the signaling cycle. While an individual signaling cycle is simply an element of a large signaling network, understanding its response is an essential first step in characterizing the response of more elaborate signaling networks to an external stimulus [1, 2].

Each cycle consists of a substrate protein that can be in one of two states: active (e.g., phosphorylated) or inactive (e.g., dephosphorylated), see Fig. 1. The protein is activated by a protein kinase that catalyzes a phosphorylation reaction. The protein gets inactivated by a second enzymatic reaction catalyzed by a phosphatase. The activity/concentration of the kinase can be considered as an input of the cycle. The response of the cycle is the level of phosphorylated substrate protein that is not bound to the phosphatase and can thus interact with any downstream components of the signaling pathway.

Signaling cycles can also require multiple phosphorylations for activation. Furthermore, cycles of phosphorylation are frequently organized into cascades where the activated substrate protein serves as a kinase for the next cycle. Activation of the first kinase in a cascade can be triggered by a receptor that has received a specific stimulus (ligand, photon, dimerization, etc.). In addition, feedback processes may be present. Furthermore, reactions may involve shuttling participating molecules between different cellular compartments, and other spatial effects. The dynamics of signaling cascades have been the subject of active research using modeling and experiments. Theoretical and computational studies of eukaryotic signaling cascades span a broad range of questions such as those concerning the dynamics of the EGFR [3] or apoptosis signaling pathways [4], the propagation of noise and stochastic fluctuations [5, 6, 7], the role of feedbacks [8, 9, 10, 11] and scaffolding proteins [12, 13], the contribution of receptor trafficking [14] and spatial effects [15, 16, 10], the origin of bistability [17, 18, 19] and oscillations [6, 20], and the consequences of multiple phosphorylations [6, 21, 22. 20, 23, 24, 25, 26].

In this paper, our focus will be on the statics and dynamics of the basic singly modified signaling cycle, with no spatial effects. The seminal contribution Goldbeter and Koshland considered the steady-state response of this basic cycle and demonstrated that, under appropriate conditions, the response can be in a highly sigmoidal ultrasensitive regime, or in a hyperbolic regime [27] (see below). Most modeling studies have assumed that all signaling cycles operate in the ultrasensitive regime; a few studies have also considered the hyperbolic regime [28, 29]. Here we demonstrate that there are actually four major regimes, with the ultrasensitive and hyperbolic regimes being two of them.

Several previous studies that treat signaling cycles as modules have focused on the steady-state response to a constant input, largely ignoring responses to time-varying stimuli (see e.g. [27,30,22]. A study of Detwiler *et al.* [28] considered the dynamic response of the cycle in the hyperbolic regime (when both forward and backward reactions are first-order), and found low-pass filtering behavior. We also recently examined the dynamic response of these two regimes and compared them in their robustness to intrinsic and extrinsic noise [31].

Here we systematically consider both the steady-state response, and the dynamic response to time-varying stimuli. To model the enzymatic reactions in the signaling cycle, we use the total quasi-steady-state approximation (tQSSA) [32]. The tQSSA is valid more generally than the Michaelis-Menten rate law, which assumes the enzyme to be present in much smaller concentration than its substrate, an assumption that is not generally valid in signaling pathways. We then use our model to examine possible regimes of the cycle, and identify two new steady-state regimes, for a total of four different behaviors, each being potentially useful in different signaling applications. Although these four regimes are defined at extreme parameter values, we numerically show that in fact together they cover almost the full parameter space. We obtain analytic approximations to the steady-state characteristics of each of the four regimes, and refine the conditions under which the two regimes identified by Goldbeter and Koshland are in fact achieved.

To obtain a fuller picture of the signaling cycle and its function, we then analyze its response to time-varying kinase activity. We demonstrate analytically that the intrinsic architecture of the cycles makes them act - in all four regimes - as tunable low-pass filters for small enough time-varying deviations of the kinase activity from baseline levels. Numerical simulations show that these analytical results continue to hold quite well even for bigger deviations from baseline level.

The four different regimes of the signaling cycle make it a versatile element, able to perform various signaling functions, while its low-pass filtering enables it to operate in noisy environments. These properties may help explain why signaling cycles are so ubiquitous in cell signaling.

# Results

## Model

The signaling cycle is modeled by two enzymatic reactions, as illustrated in Fig. 1: A forward enzymatic reaction catalyzed by kinases (enzyme 1, $E_1$) produces active proteins ($A$) from the inactive ones ($I$), and a backward reaction catalyzed by phosphatases (enzyme 2, $E_2$) de-activates active proteins:

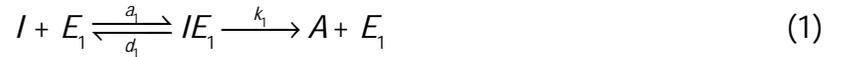
$$I + E_1 \underset{d_1}{\overset{a_1}{\rightleftharpoons}} IE_1 \xrightarrow{k_1} A + E_1 \tag{1}$$

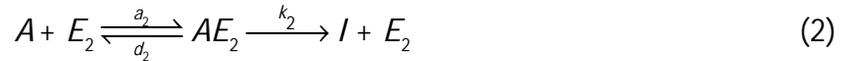
$$A + E_2 \underset{d_2}{\overset{a_2}{\rightleftharpoons}} AE_2 \xrightarrow{k_2} I + E_2 \tag{2}$$

Here $a_1$ ($d_1$) and $a_2$ ($d_2$) are substrate-enzyme association (dissociation) rates, and $k_1$ ($k_2$) is the catalytic rate of the forward (backward) enzymatic reaction. For notational convenience, we shall use the same symbol to denote a chemical species as well as its concentration. The input to the cycle is the total concentration of the active kinase, $\overline{E_1} = E_1 + IE_1$, while the output is the concentration of the free (i.e., not bound to phosphatase) active protein $A$.

While such systems are usually studied using Briggs-Haldane or Michaelis-Menten (MM) approximations (see [33, 34], both can be inapplicable as they assume much lower concentration of the enzyme than of the substrate. In fact, substrates and enzymes of MAPK pathways are usually present at comparable concentrations in *S. cerevisiae* and *Xenopus* oocyte cells (as reported in [30] and consistent with data from the library of GFP-tagged proteins [35]).

Instead, we rely on the total quasi-steady-state approximation (tQSSA)[32, 36, 37, 38] (see Methods) to obtain the following equation for the concentration of the total active protein, $\overline{A} = A + AE_2$:

$$\frac{d\overline{A}(t)}{dt} = k_1 \frac{\overline{E_1}(\overline{S} - \overline{A}(t))}{K_1 + \overline{E_1} + \overline{S} - \overline{A}(t)} - k_2 \frac{\overline{E_2}\,\overline{A}(t)}{K_2 + \overline{E_2} + \overline{A}(t)}. \tag{3}$$

Here $X$ denotes the concentration of an unbound chemical species and $\overline{X}$ denotes the total concentration of bound and unbound forms; $\overline{S}$ stands for the total concentration of substrate protein (in both active and inactive forms); and $K_1 = \frac{k_1 + d_1}{a_1}$ and $K_2 = \frac{k_2 + d_2}{a_2}$ are the MM constants for the kinase and the phosphatase, respectively. We have written $\overline{A}(t)$ explicitly with its time argument $t$ to emphasize that it is a dynamic variable; however, for notational simplicity, we will omit the time argument in the rest of the paper and simply write $\overline{A}$. The quantities $\overline{E_1}, \overline{E_2}$ and $\overline{S}$ are constant here (although later in the paper we consider the dynamic response to small variations in $\overline{E_1}$). Even though the above equation is written in terms of $\overline{A}$, the free active protein concentration $A$, which is of primary interest, is simply recovered through the expression $A = \frac{K_2 + \overline{A}}{K_2 + \overline{E_2} + \overline{A}} \overline{A}$ (see Text S1).

Equation (3) shows the dependence of the rate of production of the active protein on the number of kinases through the first term (phosphorylation), and on the number of phosphatases through the second term (dephosphorylation). In particular, when the total amounts of both kinase and phosphatase are small ($\overline{E_1} \ll K_1 + \overline{S} - \overline{A}$ and $\overline{E_2} \ll K_2 + \overline{A}$), the two terms in Equation (3) reduce to the standard MM rates for the forward and backward enzymatic reactions of the cycle. The tQSSA has also been recently proposed and applied by Ciliberto *et al.* in [39] to model networks of coupled enzymatic reactions, including interconnections of phosphorylation cycles; their reduced tQSSA representation accurately reproduces behavior predicted by detailed mass action kinetic models.

Our key equation (3) simplifies for extreme combinations of parameter values (i.e., regimes) that are still of potential biological interest. This equation allows us to analytically examine (a) the possible cycle regimes of the system in steady state, and (b) the dynamic response of the system to time-varying inputs (time-varying activation of the kinase). The numerical results we present here are not constrained by the quality of the approximation since they are based on direct simulation of the mass action kinetics equation for the full system of reactions of Equations 1 and 2 (see Methods).

## Four Regimes of the Signaling Cycle

Each enzymatic reaction can be in one of two qualitatively different regimes: a saturated one where almost all the enzyme is bound to its substrate, and an unsaturated one [40, 41]. The regime of the reaction depends on the relative concentrations of a substrate and the enzyme ($E$), and on the MM constant ($K$) of the enzymatic reaction. The unsaturated (first-order) regime, where the rate of reaction is linearly proportional to the substrate concentration, occurs when the substrate is much less abundant than the sum of the MM constant of the reaction and the enzyme concentration (e.g., for the second reaction, $K_2 + \overline{E_2} \gg \overline{A}$). In the saturated (zero-order) regime, the rate of reaction is almost independent of the substrate concentration and is proportional to the enzyme concentration. This occurs when the substrate is much more abundant than the sum of enzyme concentration and its MM constant (e.g., for the second reaction, $K_2 + \overline{E_2} \ll \overline{A}$).

Since the signaling cycle is built of two enzymatic reactions, it can exhibit four regimes of signaling (see Fig. 2), corresponding to the two regimes of each reaction. The conditions for each of the four regimes are summarized in Table 1. The steady-state behavior of two of the four regimes (when the kinase and the phosphatase are either both saturated or both unsaturated, referred to as ultrasensitive and hyperbolic, respectively) has been characterized earlier by Goldbeter and Koshland [27]. Using tQSSA, we are able to refine the range of parameter values for which these behaviors hold. The other two regimes have not been identified before, to the best of our knowledge.

## Steady-State Response

### Hyperbolic (unsaturated kinase, unsaturated phosphatase)

In this regime, the cycle exhibits a hyperbolic steady-state response that saturates at the value provided in Table 2 (see Fig. 2A). Using the tQSSA, we find that the hyperbolic regime requires weaker conditions than previously thought ($K_2 + \overline{E_2} \gg \overline{A}$ and $K_1 + \overline{E_1} \gg \overline{S} - \overline{A}$ instead of $K_2 \gg \overline{A}$ and $K_1 \gg \overline{S} - \overline{A}$).

Our recent study [31] suggests that the hyperbolic regime is much more robust to fluctuations and to cell-to-cell variability in kinase and phosphatase concentrations than the ultrasensitive regime, which requires fine-tuning of the threshold level. The hyperbolic regime transmits signals in a broad range of amplitudes, requiring no tuning of cycle parameters [31].

### Signal-Transducing (saturated kinase, unsaturated phosphatase)

We refer to this new regime as signal-transducing because, as discussed below, it is ideal for transmitting time-varying signals without distortion, while attenuating higher frequency noise. Here we only point out that its steady-state response is linear with a slope (gain) of $k_1 / \omega_2$, where $\omega_2$ is referred to as the effective phosphatase frequency (see Table 2 and section on Dynamic Response below), until it reaches saturation (Fig. 2B, Table 2). Interestingly, the total amount of substrate protein only affects the saturation level and not the slope. Therefore, away from saturation the cycle's activity is independent of the total susbtrate protein level $\overline{S}$. Having a linear steady-state response, a property unique to this regime, is potentially desirable for signaling that involves graded stimuli. Available biochemical data and *in vivo* measurements argue for the possibility of this regime being present as a component in cell signaling cascades (see Discussion).

### Threshold-Hyperbolic (unsaturated kinase, saturated phosphatase)

In this new regime, the output below a given input threshold is zero, and then increases hyperbolically until it reaches its saturation level (approximated by the same expression as the saturation level of an ultrasensitive regime). Figure 2C shows the steady-state response of such a cycle.

### Ultrasensitive (saturated kinase, saturated phosphatase)

The output in this regime is close to zero for inputs below a threshold, and increases rapidly to a saturation value, consistent with the results obtained in [27] using the MM approximation. Such highly sigmoidal behavior effectively quantizes the signal (see Fig. 2D). This regime was termed ultrasensitive because, when the input is close to the threshold, small input changes result in large changes of the steady-state output. Interestingly, cells may adjust the threshold of this cycle by changes in phosphatase level, $\overline{E_2}$.

The MM approximation fails, however, when the amount of enzyme becomes comparable to that of its substrate. Using the tQSSA we are able to refine the range of parameter values required for ultrasensitive signaling. The criteria for ultrasensitivity obtained from the MM model [27], namely $K_2 \ll \overline{A}$ and $K_1 \ll \overline{S} - \overline{A}$, are actually not sufficient conditions for the cycle to be ultrasensitive; instead we need $K_2 + \overline{E_2} \ll \overline{A}$ and $K_1 + \overline{E_1} \ll \overline{S} - \overline{A}$. When the enzyme concentrations become comparable to those of their substrates, there is no ultrasensitivity, as noted recently by Bluthgen *et al.* [42] by more complicated arguments.

In summary, we have demonstrated that a signaling cycle can operate in four regimes that have qualitatively different steady-state responses to kinase activation. Of the newly identified regimes, the signal-transducing regime is a good candidate for sensing stimuli, when a graded and undistorted response is required. Depending on the slope of its response, which is controlled by parameters of the cycle and can be easily adjusted by the cell to a required level, the input signal may be amplified or diminished. We consider factors influencing the choice of the regime for natural signaling cycles in different cellular processes in the Discussion.

The four regimes we consider, although obtained only at extreme parameter values, are actually quite descriptive of the system for a wide range of parameters, and naturally partition the space of possible steady-state behaviors of the signaling cycle into quadrants, as shown Fig. 3. This figure shows the relative error between the steady-state characteristic of each of the four regimes and that of Eq. 3 for a wide range of kinase and phosphatase MM constants (see Text S4). It reveals that the regime approximations are quite good at a wide range of values of MM constant (for example, the region with a relative error of less than 10% for each regime covers almost a full quadrant in the plots), and not only when the MM constants take the very large or very small values required in the regime definitions. This demonstrates that these four regimes, though defined by extreme values of system parameters, actually encompass the full space of cycle behaviors.

Understanding the steady-state response of the cycle is informative, but is only part of the story; signaling cycles do not necessarily transmit steady inputs but rather deal with time-dependent signals that reflect changing environmental conditions.

## Dynamic Response

Signaling cascades in the cell are activated by receptors, which in turn get activated by ligand binding and inactivated by internalization and other mechanisms. All of these mechanisms produce time-varying signals, and are subjected to noise (i.e., rapid and stochastic fluctuations) due to small numbers of molecules, diffusion, and other effects. How can a cell extract a time-varying signal from noisy stimuli?

**Response to signals of various frequencies: low-pass filtering**

To address this question, we first study the response of the four regimes to time-varying stimuli. A high-frequency signal is a proxy for the noise in the signal, so understanding how the cycle responds to high frequencies is essential for understanding its response to noise.

We studied the cycle's response to oscillating kinase levels at different frequencies and amplitudes: $\overline{E_1}(t) = E_0(1 + a\sin\omega t)$. This is not to say that sinusoidal inputs need be biologically relevant, but systematically understanding the response to such inputs gives one intuition about the response to more general inputs. Furthermore, for small enough input variations around some background baseline level, the cycle's behavior is, to a first-order approximation, linear and governed by time-independent parameters; in this situation the response to sinusoids determines the response to arbitrary inputs. In fact, in the signal-transducing regime the dynamic response of the cycle (just as its static response) is linear for all non-saturating inputs, without the restriction to small variations. If all cycles in a signaling pathway are in a linear regime, then analysis of the overall behavior is ammenable to standard and very effective methods.

Figure 4 shows the amplitude $O$ of the variations in the output (normalized by the steady-state saturation value of the cycle), obtained by numerical simulation for three values of $a$, and as a function of input frequency $\omega$. Invariably, the response is flat and high at low frequencies, but starts to decrease after a particular frequency is reached. These results are very well described in the case of the smallest $a$ (corresponding to 11% deviations) by the expression obtained analytically using small-signal approximations (see Text S4):

$$O = \left(\frac{g}{\sqrt{\omega^2 + \omega_c^2}}\right) E_0 a, \qquad (4)$$

where $E_0$ is the background kinase level, and where the gain $g$ and the cutoff frequency $\omega_c$ are functions of the cycle parameters that are different for the four regimes (see Table 3). The analytical approximation continues to hold quite well even for larger values of $a$, the deviation amplitude (up to 91% of the baseline for the results in Figure 4). For frequencies much smaller than the cutoff, the amplitude of the output variations is constant and proportional to the ratio of gain to cutoff frequency. For frequencies above the cutoff, the output variations have an amplitude that decays as $1/\omega$. Figure S1 presents more detailed results on the variation of $O$ as a function of both $a$ and $\omega$, again obtained by numerical simulations (see Text S6 for a description of Figure S1).

An essential property of this signaling low-pass filter is that the cutoff frequency $\omega_c$ can be easily adjusted by varying enzymatic parameters and concentrations of the kinase and the phosphatase. Although all four regimes act as low-pass filters, their cut-off frequencies $\omega_c$ and gains $g$ depend differently on the cycle parameters (see Table 3 and Fig. 4).

Importantly, for the two newly characterized regimes (the signal-transducing and threshold-hyperbolic), the gain and the cutoff frequency can be adjusted independently, thus allowing greater flexibility to the signaling requirements of individual signaling pathways (Table 3 and Fig. 4). Furthermore, the dynamics of the signal transducing regime again do not depend on total substrate protein levels $\overline{S}$. The gain and the cut-off frequency for three of the regimes are independent of the input parameters $a$ and $\omega$; the exception is the cutoff frequency for the ultrasensitive regime, which depends on the input amplitude.

It is easy to understand the origin of the low-pass filtering behavior. First consider a cycle subjected to a slowly varying input (Fig. 5): If the input changes so slowly that the cycle has enough time to reach its steady-state level before the kinase level changes by a significant amount, the cycle simply tracks the kinase level as a function of time through its steady-state response curve, characteristic for its operational regime. Now consider a rapidly changing input. Since the kinase level changes faster, the cycle has less time to adjust to its steady state corresponding to the new value of the input before the kinase level changes again. Thus the output will not be able to reach its full amplitude before the kinase levels change again in the opposite direction, and the amplitude of the output is thus decreased (see Fig. 5). As the signal changes faster and faster, the amplitude of the output will decrease, until the kinase levels vary so fast that the cycle simply does not respond.

The response of the cycle thus depends on the two time-scales: the duration of the stimulus $\tau = 1/\omega$ and the intrinsic switching time of the cycle $\tau_c = 1/\omega_c$. If the stimulation is longer than the switching time, $\tau \gg \tau_c$, then the cycle will adjust its response by $2aE_0 g/\omega_c$. On the other hand, a shorter, transient stimulus $\tau \ll \tau_c$ is not likely to activate the cascade.

Interestingly, ligands activate a kinase by binding to it. The results here imply that weak ligands binding for a time interval shorter than $\tau_c$ are unlikely to produce any down-stream activation of the pathway, while those that stay bound longer than $\tau_c$ activate the pathway. Low-pass filtering can thus perhaps make a signaling cascade more selective to higher-affinity ligands.

**Response to a noisy signal**

Importantly, low-frequency inputs are proxies for longer input activation, while high-frequency inputs are proxies for short, transient activations of the cascade and for high-frequency noise. Because of low-pass filtering, cycles respond to noise less than to signals, and as the noise shifts to higher frequencies, the cycle responds to it less. Figure 6 makes the point more precisely: it shows the response of the cycle to a slowly varying signal buried in noise, and demonstrates that the noise is filtered out and the signal is revealed.

In summary, analysis of dynamic response demonstrates that (i) the cycle acts as a low-pass filter in all four regimes; (ii) the cutoff frequency and the gain of signaling can be adjusted by the cell to achieve better performance (independently of each other in the case of the signal-transducing and the threshold-hyperbolic cycles); and (iii) low-pass filtering makes signaling cascades insensitive to noise and transient activations. Below we discuss some biological implications of these findings.

# Discussion

Significant effort has been put in the elucidation and characterization of signaling cascades and pathways (see e.g. [43, 44, 2, 16] for reviews). When put together, these pathways form an intricate network of cell signaling, where each node in the network corresponds to a different chemical species. Because of the complexity of the network, it is natural to split it into interconnected modules (sets of nodes whose output depends only on its input and not on the network downstream of it) and analyze possible behaviors arising from different interconnections of modules (see e.g. [45, 29, 46]).

What constitutes a module in the network, however, is still hard to define, and significant efforts are directed at tackling this problem (e.g. [47, 48, 49, 50, 51]). What constitutes a good general representation for an arbitrary module in the network is also an open question. Other efforts have been aimed at understanding properties of the network as a whole, such as identifying the number of equilibrium states (e.g. [52, 53]).

Using a deterministic model, we have attempted to provide a systems-level input/output understanding of the signaling cycle, ubiquitous in signaling pathways. After identifying four parameter regimes (two of them not reported before, to our knowledge), their steady-state and dynamic behaviors were analyzed and numerically verified. The results indicate that cycles act as low-pass filters, and that each regime may be useful under different circumstances. Given the values for cycle parameters, one can use our results to determine the regime in which the cycle operates. Unfortunately, the scarcity of parameter values makes it hard to assess which of these regimes is more widely present in signaling pathways. The low-pass filtering behavior of the cycle demonstrates that inputs of the same magnitude but changing at different speeds may produce very different outputs, which argues in favor of studying the dynamical properties of signaling pathways.

All physical systems stop responding to fast enough inputs, but what makes the low-pass filtering behavior of the signaling cycle interesting is that it is first-order, with a single cutoff frequency, and that the cutoff frequency can be adjusted by evolution (through changes in the enzymatic catalytic rates) and by the cell (through changes in gene expression). As such, the signaling cycle is a versatile module with simple dynamics that can be easily tuned for various noise filtering needs and used to construct signaling networks with more complicated functions and dynamics.

Of the two newly identified regimes, the signal transducing one is of particular interest because it appears ideal to transmit time-varying intracellular signals without distortion, while filtering out high-frequency noise in the input. Furthermore, because it is linear, it opens the possibility that at least parts of signaling pathways (those built of signal-transducing signaling cycles, or other yet unidentified linear signaling motifs) may be amenable to linear system analysis, a powerful set of tools to understand the properties of arbitrary network structures and motifs (for example, elucidating the roles of cascades, positive and negative feedbacks, etc.). If naturally occurring cycles operate in the signal-transducing regime, then analyzing networks built of these cycles becomes tractable as long as load effects can be neglected.

Can naturally occurring signaling cycles operate in this regime? While it was demonstrated that certain kinases in *S. cerevisiae* and *Xenopous* operate in saturation (with MM constant of ~5nM and substrate concentrations of ~ 30 - 100 nM for yeast [30, 54]), little is known about phosphatases. To explore the possibility that known signaling pathways operate in the signal-transducing regime, we manually collected values of MM constants from the biochemical literature. Then we used data for intracellular protein concentrations measured using GFP-tagged proteins [35]. Phosphatases seem to have a broad specificity, with a relatively wide range of MM constants (e.g., 5 to 90 $\mu$M for the PP2C phosphatases), and appear to be present in large concentrations (e.g., [Ptc1] $\approx$ 1520 molecules per cell, so $\overline{E}_2 \approx 0.025 \mu M$, while [Ptc2-3] $\approx$ 15000, so $\overline{E}_2 \approx 0.025 \mu M$, assuming a yeast cell volume of 0.1 pl [30]). Data on singly-phosphorylated substrates is hard to find, but for a rough indication consider the doubly phosphorylated protein Pbs2 of *S. cerevisiae* as an example. Pbs2 is measured to have about 2000 molecules per cell so that $\overline{S} \approx 0.03 \mu M$ = 30nM. If singly phosphorylated proteins were characterized by similar numbers, then their phosphatases could potentially be unsaturated, since $\overline{A} < \overline{S} \ll K_2 + \overline{E}_2$. In contrast, kinases that act on Pbs2 are present at lower concentrations (e.g., [Ste11] = 736, [Ssk2] = 217, and [Ssk22] = 57 molecules per cell, or $\overline{E}_1 \approx$ 1-3 nM). Such concentrations are consistent with kinases operating in saturation, since $\overline{E}_1 + K_1 < \overline{S}$ (assuming $K_1$ is in the same range as those measured for Ste7, $K_1 \sim$ 5 nM). Taken together, these numbers suggest that the possibility of a signaling cycle operating in the signal-transducing regime.

Different signaling cycles, however, may be operating in different regimes, raising two questions: First, which regime is chosen by the cell for a cycle in a particular position in a network for a specific signaling application? Second, what are the advantages and disadvantages of each such regime?

To answer the first question, one approach is to determine *in vivo* concentrations and MM constants of involved enzymes. Unfortunately, these data are often unavailable or scattered through publications in the biochemical literature. The applicability of MM constants measured *in vitro* is also questionable. An alternative experimental approach to establishing what regime a cycle operates in would be to obtain its steady-state response curve and determine which of our four cases it corresponds to. Similarly, one could experimentally obtain the response of the cycle to stimuli of various frequencies and use our dynamic characterization to infer the operating regime. One may furthermore be able to estimate some of the biochemical parameters and concentrations of the participating molecules from these experimental response characteristics. The success of such measurements depends on and hence is limited by the availability of *in vivo* single cell probes for the phosphorylation state of a particular protein.

The second question, on advantages and disadvantages of each regime, can be addressed by systematic analysis of cycle properties: steady-state and dynamic response, robustness to fluctuations, etc. By matching these characteristics against the requirements of a particular signaling system, one can suggest the optimal regime for each signaling application. For example, one can think that signaling in retina cells shall be fast and graded, depending on the intensity of adsorbed light. Similarly, gradient sensing in motile cells has to provide graded responses on the time-scales required to change direction of motion. On the other hand, signaling of cell fate determining stimuli and signaling involved in various developmental processes may require an ultrasensitive ("on/off") response, while imposing much softer constraints on the time it takes to switch the system from off to on state (hours instead of the milliseconds needed in light-sensing). The performance of the signaling regimes in the context of cascades and feedbacks is also important for understanding the rules that govern the choice of a regime for each cycle.

For cycles in signaling applications involving all-or-none decisions, such as differentiation, apoptosis, or the cell cycle, it has been argued that ultrasensitive cycles may be useful as they effectively generate a discrete output that is either high or low [24]. When such a cycle is tuned appropriately (such that in the presence of the background input it is close to its threshold) [31], it is the best cycle at recovering time-dependent signals buried in noise, because its gain for low-frequency inputs is the highest among the regimes. Therefore an ultrasensitive cycle is desirable when the input signals are extremely noisy, and/or have to achieve binary level outputs.

A signal-transducing cycle, on the other hand, is the best choice to transmit time-dependent signals without distortion because its output is approximately a scaled but otherwise undistorted copy of low-frequency input signals, while noisy input components are filtered out. It is the only cycle that does not distort the input. What the other two regimes might be best at is not clear. The threshold-hyperbolic cycle, however, may prove useful in situations when no activation is desirable below a given input strength, and when a graded response is desired for inputs above this threshold.

We here considered the effect of temporal noise in kinase levels on the response of the signaling cycle. A more detailed model should also take into account the intrinsic noise coming from the cycle itself, since it consists of chemical reactions where the number of molecules per species is small, and thus a deterministic model based on mass action kinetics may be inadequate. For example, although the deterministic cycle is known to have a single steady-state solution, Samoilov *et al.* (see [6]) found that treating the cycle stochastically can give rise to a bimodal distribution for the phosphorylated protein. The "mass fluctuation kinetics" approach described in [55] may be useful in this regard, see also [56, 57]. Other sources of noise that should also be taken into account are fluctuations in molecule numbers from cell to cell, as has been well documented for gene levels (see [5, 58, 59], for example). Lastly, some of the species of the cycle may be found only in the cellular membrane rather than in the cytoplasm, or may be localized within specific cellular compartments, or may move about the cell by diffusion or active transport in an activity-dependent manner (e.g., the yeast protein HOG1 that dwells in the cytoplasm unless doubly phosphorylated, when it translocates into the cell nucleus). The consequences of these spatial effects need to be understood (see [16] for a recent review).

Achieving a detailed understanding of signaling pathways is an important problem, but is highly limited by the lack of experimental data with enough resolution to support modeling efforts. Nevertheless, having coarse-grained functional characterizations of the possible operating regimes of constituent cycles may permit system level modeling of networks built of such cycles, despite uncertainties and variations in underlying biochemical parameters and molecular concentrations. Perhaps identifying and analyzing other relevant modules of biological networks, as we have done here for a signaling cycle, will shed some light on their behavior. Similar explorations could be done, for example, on signaling cycles that require multiple phosphorylation events to become active, or on G-protein coupled receptors.

Although characterization of the component modules of a biological network is a necessary and important step towards understanding network operation, it should be kept in mind that the behavior of the network will undoubtedly be considerably richer than that of the individual modules.

The SwissProt [60] reference numbers for the genes Ptc1, Ptc2, Ptc3, Pbs2, Ste7, Ste11, Ssk2 and Ssk22, mentioned in this paper, are respectively P35182, P39966, P34221, P08018, P06784, P23561, P53599, P25390.

## Methods

All analytical expressions were obtained starting from Equation 3, the tQSSA approximation of the cycle, the derivation of which is discussed in Text S1. The full mass action kinetics (MAK) description of the system (again, see Text S1) was analyzed numerically to obtain the data used in all the plots. Therefore, although the analytical expressions in this paper depend on the validity of the tQSSA, the general results do not as they have been numerically verified on the full system.

The cycle equation (Equation 3) corresponding to each regime is described in Text S2. These equations were then used to obtain the steady-state expressions in Table 2; see Text S3. The expression for the amplitude of the response to sinusoidal inputs (Equation 4) was obtained from a small-signal approximation of Equation 3, as described in Text S4. There we also outline the method to obtain the expressions in Table 3.

All numerical analysis was done in Matlab and, unless explicitly mentioned here, is based on the full MAK description of the cycle. The data in Figure 2 was obtained by setting the derivatives of the MAK model to zero and solving the resulting algebraic relations numerically. The data in Figure 3 is the only one based on the tQSSA, and is described in Text S5. Figures 4, 5, and S1 were obtained by numerically integrating the MAK equations for the given inputs using the stiff differential equation solver from Matlab *ODE23s.* Finally, the data in Figure 6 was obtained by numerically integrating the MAK equations using the Runge-Kutta algorithm on inputs of the form $E_0(1+a\sin\omega t_i+\eta(0,1))$ (where $t_i$ is any time point in the numerical integration and $\eta(0,1)$ is a normal random variable with unit variance and zero mean). All the code is available upon request.

For all the dynamic simulations, the steady-state level of the input for the four cycles was chosen such that the steady-state output was about half-way to saturation, to allow the cycles to respond as much as possible. Choosing other steady-state values where the slope of the steady state response curve is small would lead to little response. Particular care has to be taken with in the ultrasensitive cycle, which has a very small range of inputs where its slope is non-zero, implying that this cycle needs to be finely tuned for it to transmit dynamic information (see Text S4).

## Acknowledgments

C.A.G.-U. gratefully acknowledges the support of an MIT-Merck Graduate Fellowship, and the MIT EECS/Whitehead/Broad Training Program in Computational Biology (Grant 5 R90 DK071511-01 from the NIH). LM acknowledges the support of the National Center for Biomedical Computing *i2b2*. We also thank Rami Tzafriri for suggesting the use of the tQSSA to reduce the mass action kinetics system.

# References

[1] Rao CV, Arkin AP (2001) Control motifs for intracellular regulatory networks. Annu Rev Biomed Eng 3: 391-419.

[2] Sauro HM, Kholodenko BN (2004) Quantitative analysis of signaling networks. Progress in biophysics and molecular biology 86: 5-43.

[3] Wiley HS, Shvartsman SY, Lauffenburger DA (2003) Computational modeling of the egf-receptor system: a paradigm for systems biology. Trends Cell Biol 13: 43-50.

[4] Li C, Ge QW, Nakata M, Matsuno H, Miyano S (2007) Modelling and simulation of signal transductions in an apoptosis pathway by using timed petri nets. J Biosci 32: 113-127.

[5] Thattai M, van Oudenaarden A (2004) Stochastic gene expression in fluctuating environments. Genetics 167: 523-530.

[6] Samoilov M, Plyasunov S, Arkin AP (2005) Stochastic amplification and signaling in enzymatic futile cycles through noise-induced bistability with oscillations. Proc Natl Acad Sci USA 102: 2310-2315.

[7] Bialek W, Setayeshgar S (2005) Physical limits to biochemical signaling. Proc Natl Acad Sci USA 102: 10040-10045.

[8] Brandman O, Ferrell JEJ, Li R, Meyer T (2005) Interlinked fast and slow positive feedback loops drive reliable cell decisions. Science 310: 496-498.

[9] Legewie S, Bluthgen N, Herzel H (2006) Mathematical modeling identifies inhibitors of apoptosis as mediators of positive feedback and bistability. PLoS Comput Biol 2: e120.

[10] Levchenko A, Iglesias PA (2002) Models of eukaryotic gradient sensing: application to chemotaxis of amoebae and neutrophils. Biophys J 82: 50-63.

[11] Kholodenko BN (2000) Negative feedback and ultrasensitivity can bring about oscillations in the mitogen-activated protein kinase cascades. European journal of biochemistry / FEBS 267: 1583-1588.

[12] Levchenko A, Bruck J, Sternberg PW (2000) Scaffold proteins may biphasically affect the levels of mitogen-activated protein kinase signaling and reduce its threshold properties. Proc Natl Acad Sci USA 97: 5818-5823.

[13] Borisov NM, Markevich NI, Hoek JB, Kholodenko BN (2005) Signaling through receptors and scaffolds: independent interactions reduce combinatorial complexity. Biophys J 89: 951-966.

[14] Vilar JMG, Jansen R, Sander C (2006) Signal processing in the tgf-beta superfamily ligand-receptor network. PLoS Comput Biol 2: e3.

[15] Ander M, Beltrao P, Di Ventura B, Ferkinghoff-Borg J, Foglierini M, et al. (2004) Smartcell, a framework to simulate cellular processes that combines stochastic approximation with diffusion and

localisation: analysis of simple networks.  Syst Biol (Stevenage) 1: 129-138.

[16] Kholodenko BN (2006) Cell-signalling dynamics in time and space.  Nature Reviews Molecular Cell Biology 7: 165-166.

[17] Markevich NI, Hoek JB, Kholodenko BN (2004) Signaling switches and bistability arising from multisite phosphorylation in protein kinase cascades.  J Cell Biol 164: 353-359.

[18] Paliwal S, Iglesias PA, Campbell K, Hilioti Z, Groisman A, et al. (2007) Mapk-mediated bimodal gene expression and adaptive gradient sensing in yeast.  Nature 446: 46-51.

[19] Legewie S, Bluthgen N, Schafer R, Herzel H (2005) Ultrasensitization: switch-like regulation of cellular signaling by transcriptional induction.  PLoS Comput Biol 1: e54.

[20] Chickarmane V, Kholodenko BN, Sauro HM (2007) Oscillatory dynamics arising from competitive inhibition and multisite phosphorylation.  J Theor Biol 244: 68-76.

[21] Swain PS, Siggia ED (2002) The role of proofreading in signal transduction specificity.  Biophys J 82: 2928-2933.

[22] Gunawardena J (2005) Multisite protein phosphorylation makes a good threshold but can be a poor switch.  Proc Natl Acad Sci USA 102: 14617-14622.

[23] Ortega F, Garces JL, Mas F, Kholodenko BN, Cascante M (2006) Bistability from double phosphorylation in signal transduction. kinetic and structural requirements.  FEBS J 273: 3915-3926.

[24] Ferrell Jr JE, Machleder EM (1998) The biochemical basis of an all-or-none cell fate switch in Xenopus oocytes.  Science 280: 895-898.

[25] Ferrell Jr JE (1999) Building a cellular switch: more lessons from a good egg.  BioEssays : news and reviews in molecular, cellular and developmental biology 21: 866-870.

[26] Wang L, Sontag ED (2007) A remark on the number of steady states in a multiple futile cycle.  Q-bio arXiv:0704.0036v1

[27] Goldbeter A, Koshland Jr DE (1981) An amplified sensitivity arising from covalent modification in biological systems.  Proc Natl Acad Sci USA 78: 6840-6844.

[28] Detwiler PB, Ramanathan S, Sengupta A, Shraiman BI (2000) Engineering aspects of enzymatic signal transduction: photoreceptors in the retina.  Biophys J 79: 2801-2817.

[29] Heinrich R, Neel BG, Rapoport TA (2002) Mathematical models of protein kinase signal transduction.  Molecular Cell 9: 957-960.

[30] Ferrell Jr JE (1996) Tripping the switch fantastic: how a protein kinase cascade can convert graded inputs into switch-like outputs.  Trends in Biochemical Sciences 21: 460-466.

[31] Levine J, Kueh HY, Mirny LA (2007) Intrinsic Fluctuations, Robustness and Tunability in

Signaling. Biophys J 92:4473-81.

[32] Tzafriri AR (2003) Michaelis-Menten kinetics at high enzyme concentrations. Bulletin of mathematical biology 65: 1111-1119.

[33] Briggs GE, Haldane JB (1925) A Note on the Kinetics of Enzyme Action. The Biochemical Journal 19: 338-339.

[34] Michaelis L, Menten M (1913) Die Kinetik der Invertinwirkung. The Biochemical Journal Z 49: 333-369.

[35] Ghaemmaghami S, Huh WK, Bower K, Howson RW, Belle A, et al. (2003) Global analysis of protein expression in yeast. Nature 425: 737-741.

[36] Tzafriri AR, Edelman ER (2004) The total quasi-steady-state approximation is valid for reversible enzyme kinetics. Journal of Theoretical Biology 226: 303-303.

[37] Tzafriri AR, Edelman ER (2005) On the validity of the quasi-steady state approximation of bimolecular reactions in solution. Journal of Theoretical Biology 233: 343-350.

[38] Schnell S, Maini PK (2000) Enzyme kinetics at high enzyme concentration. Bulletin of mathematical biology 62: 483-489.

[39] Ciliberto A, Capuani F, Tyson JJ (2007) Modeling networks of coupled enzymatic reactions using the total quasi-steady state approximation. PLoS Comput Biol 3: e45.

[40] Fersht A (1998) Structure and Mechanism in Protein Science: A Guide to Enzyme Catalysis and Protein Folding. NY: W. H. Freeman.

[41] Voet D, Voet J (1995) Biochemistry. 2nd edition. J. Wiley & Sons, New York.

[42] Bluthgen N, Bruggeman FJ, Legewie S, Herzel H, Westerhoff HV, et al. (2006) Effects of sequestration on signal transduction cascades. FEBS J 273: 895-906.

[43] The Alliance for Cellular Signaling (2002). http://www.afcs.org.

[44] Steffen M, Petti A, Aach J, D'haeseleer P, Church G (2002) Automated modelling of signal transduction networks. BMC bioinformatics 3: 34.

[45] Hartwell LH, Hopfield JJ, Leibler S, Murray AW (1999) From molecular to modular cell biology. Nature 402: 47-52.

[46] Bruggeman FJ, Westerhoff HV, Hoek JB, Kholodenko BN (2002) Modular response analysis of cellular regulatory networks. Journal of Theoretical Biology 218: 507-510.

[47] Hofmann KP, Spahn CM, Heinrich R, Heinemann U (2006) Building functional modules from molecular interactions. Trends in Biochemical Sciences 31: 497-508.

[48] Spirin V, Gelfand MS, Mironov AA, Mirny LA (2006) A metabolic network in the

evolutionary context: multiscale structure and modularity. Proc Natl Acad Sci USA 103:8774-8779.

[49] Spirin V, Mirny LA (2003) Protein complexes and functional modules in molecular networks. Proc Natl Acad Sci USA 100: 12123-12128.

[50] Pereira-Leal JB, Enright AJ, Ouzounis CA (2004) Detection of functional modules from protein interaction networks. Proteins 54:49-57. Evaluation Studies.

[51] Newman MEJ (2006) Modularity and community structure in networks. Proc Natl Acad Sci USA 103: 8577-8582.

[52] Angeli D, Ferrell Jr JE, Sontag ED (2004) Detection of multistability, bifurcations, and hysteresis in a large class of biological positive-feedback systems. Proc Natl Acad Sci USA 101: 1822-1827.

[53] Chaves M, Sontag ED, Dinerstein RJ (2004) Steady-states of receptor-ligand dynamics: a theoretical framework. Journal of Theoretical Biology 227: 413-418.

[54] Bardwell L, Cook JG, Chang EC, Cairns BR, Thorner J (1996) Signaling in the yeast pheromone response pathway: specific and high-affinity interaction of the mitogen-activated protein (map) kinases kss1 and fus3 with the upstream map kinase kinase ste7. Mol Cell Biol 16: 3637-3650.

[55] Gomez-Uribe CA, Verghese G (2007) Mass fluctuation kinetics: Capturing stochastic effects in systems of chemical reactions through coupled mean-variance computations. The Journal of Chemical Physics 126: 024109.

[56] Goutsias J (2007) Classical versus stochastic kinetics modeling of biochemical reaction systems. Biophys J 92: 2350-2365.

[57] Gadgil C, Lee CH, Othmer HG (2005) A stochastic analysis of first-order reaction networks. Bull Math Biol 67: 901-946.

[58] Paulsson J (2004) Summing up the noise in gene networks. Nature 427: 415-418.

[59] El-Samad H, Khammash M (2006) Regulated degradation is a mechanism for suppressing stochastic fluctuations in gene regulatory networks. Biophys J 90: 3749-3761.

[60] Swiss-Prot Database. http://expasy.org/sprot/.

[61] Huang CY, Ferrell Jr JE (1996) Ultrasensitivity in the mitogen-activated protein kinase cascade. Proc Natl Acad Sci USA 93: 10078-83.

[62] Yi TM, Huang Y, Simon MI, Doyle J (2000) Robust perfect adaptation in bacterial chemotaxis through integral feedback control. Proc Natl Acad Sci USA 97: 4649-4653.

# Tables

**Table 1** Conditions for the four cycle regimes.

| Kinase<br><br>Phosphatase | Unsaturated<br>$K_1 + \bar{E}_1 \gg \bar{S} - \bar{A}$ | Saturated<br>$K_1 + \bar{E}_1 \ll \bar{S} - \bar{A}$ |
|---|---|---|
| Unsaturated<br>$K_2 + \bar{E}_2 \gg \bar{A}$ | hyperbolic | signal transducing |
| Saturated<br>$K_2 + \bar{E}_2 \ll \bar{A}$ | threshold hyperbolic | ultrasensitive |

**Table 2** Expressions for threshold and saturation levels for steady-state regimes of the cycle. Here $\omega_2 = k_2 \dfrac{\bar{E}_2}{K_2 + \bar{E}_2}$ is the characteristic frequency of the phosphatase.

| Regime | Threshold for input ($E_1$) | Saturation level |
|---|---|---|
| hyperbolic | - | $\dfrac{k_1}{k_1 + w_2}(1 - \dfrac{w_2}{k_2})\bar{S}$ |
| signal-transducing | - | $(1 - \dfrac{\omega_2}{k_2})(\dfrac{k_1/\omega_2}{1 + k_1/\omega_2})\bar{S}$ |
| threshold-hyperbolic | $K_1 / ((k_1/k_2 - 1)(\bar{S}/\bar{E}_2 - 1))$ | $\bar{S} - \bar{E}_2(1 + \dfrac{k_2}{k_1})$ |
| ultrasensitive | $\bar{E}_2 k_2 / k_1$ | $\bar{S} - \bar{E}_2(1 + \dfrac{k_2}{k_1})$ |

**Table 3** Expressions for gain and cutoff frequency for four regimes of the cycle (in response to the input $\bar{E}_1(t) = E_0(1 + a\sin\omega t)$. Here $\omega_2$ is the characteristic frequency of the phosphatase, defined in Table 2; while $\omega_1 = k_1 \dfrac{E_0}{K_1 + E_0}$ is the characteristic frequency of the kinase.

| Regime | Gain $g$ | Cutoff Frequency $\omega_c$ |
|---|---|---|
| hyperbolic | $\dfrac{\bar{S}K_1}{E_0(K_1 + E_0)} \dfrac{\omega_2}{\omega_2 + \omega_1}(1 - \dfrac{\omega_2}{k_2})\omega_1$ | $\omega_1 + \omega_2$ |
| signal-transducing | $(1 - \dfrac{\omega_2}{k_2})k_1$ | $\omega_2$ |
| threshold-hyperbolic | $\dfrac{\bar{E}_2 K_1}{E_0(K_1 + E_0)} k_2$ | $\omega_1$ |
| ultrasensitive | $k_1$ | $2k_1 a \dfrac{E_0}{\bar{S} - \bar{E}_2(1 + k_2/k_1)}$ |

# Figure Legends

Figure 1: **Diagram of the signaling cycle.** The cycle consists of a protein that can be in an inactive ($I$) or active ($A$) form. It is activated and deactivated by two enzymatic species, termed kinase ($E_1$) and phosphatase ($E_2$), respectively. The reactions and reaction rates that describe the cycle are shown on the right.

Figure 2: **Steady state behavior of the four cycle regimes.** **A** When both enzymes are unsaturated the steady state response is hyperbolic. The parameters used for this cycle are $\overline{S} = 1000$, $a_1 = 1$, $K_1 = 10000$, $a_2 = 1$, $\overline{E_2} = 50$, $K_2 = 10000$, $k_1 = 1$ and $k_2 = 1$ where all reaction rates are in units of $1/sec$, concentrations and Michaelis constants are in $nM$, and second order reaction rates ($a_1$ and $a_2$ are in $1/nMsec$. **B** When the kinase is saturated and the phosphatase unsaturated a linear response results. The parameters here are $\overline{S} = 1000$, $a_1 = 100$, $K_1 = 10$, $a_2 = 1$, $\overline{E_2} = 50$, $K_2 = 10000$, $k_1 = 500$ and $k_2 = 10000$. **C** When the kinase is unsaturated and the phosphatase saturated a threshold-hyperbolic response results. The parameters for this cycle are $\overline{S} = 1000$, $a_1 = 100$, $K_1 = 10000$, $a_2 = 100$, $\overline{E_2} = 100$, $K_2 = 1$, $k_1 = 25$ and $k_2 = 1$. **D** When both enzymes are saturated an ultrasensitive response results. The parameters used for this cycle are $\overline{S} = 1000$, $a_1 = 100$, $K_1 = 10$, $a_2 = 100$, $\overline{E_2} = 50$, $K_2 = 10$, $k_1 = 1$ and $k_2 = 1$. The parameters for the four cycles were chosen to be comparable in magnitude to values found in the literature (see [61, 11], for example).

Figure 3: **Relative error.** Subfigures **A, B, C** and **D** respectively show the relative error between the steady-state characteristic of the hyperbolic, signal-transducing, threshold-hyperbolic and ultrasensitive regimes, and that of the tQSSA in Equation (3). To compute the error for a regime we first approximated the average squared difference between the regime's steady state and that of Equation (3), and then divided its square root by the total substrate $S_t$ (see Section E in the Protocol S1 for more details). A relative error of .1 then corresponds to an average absolute difference between the steady state characteristic of the regime and that of Equation (3) of $.1 S_t$. The figures here show that the relative error for each regime is small for a wide region of the $K_1$ versus $K_2$ space, demonstrating that the four regimes cover almost the full space. The parameters used for this cycle are the same as those in Figure 2D, excep! t $K_1$ and $K_2$ which were varied in the range of values shown in the $x$ and $y$ axes in this figure. The dashed lines enclose the regions where each regime is expected to describe the system well.

Figure 4: **Magnitude of the response of the cycle $O$ (normalized by the steady-state saturation value) versus the input frequency $\omega$, for three different input amplitudes** $a$. The traces in **A, B, C** and **D** show the response of the hyperbolic, signal transducing, threshold-hyperbolic and ultrasensitive switches, respectively, shown in Figure 2. The solid lines are the analytical approximation (Equation (4)). The dotted lines are obtained from numerical simulation of the full system.

Figure 5: **Dynamic response of the cycles to fast and slow inputs.** The cycle has a characteristic response time $\tau_c$ that is a function of its parameters (see Section Dynamic Response), and which is different for all four regimes. This plot shows the response of all four regimes to (i) a slow input that has a period equal to twice the characteristic response time of the cycle followed by (ii) a fast input with a period equal to one fifth of the cycle's response time. For clarity, time was normalized by dividing by the characteristic time of each cycle. The signal in red represents the input kinase levels (for the threshold-hyperbolic switch the input used is actually twice the red signal) while the blue traces in **A, B, C** and **D** show the response of the hyperbolic, signal transducing, threshold-hyperbolic and ultrasensitive switches, respectively, shown in Figure 2.

Figure 6: **Response of the four cycles to the input buried in noise.** The input is a sum of a slow signal (same as in Fig. 4) and a Gaussian uncorrelated noise. The resulting input signals are shown in red. The blue traces in **A, B, C** and **D** show the response of the hyperbolic, signal transducing, threshold-hyperbolic and ultrasensitive switches, respectively, as shown in Figure 2. The response shows that the cycles respond to the signal only and ignore or filter out the noise in the input. Time was normalized by the characteristic time of each cycle to facilitate comparison amongst cycles.

**Fig.1**

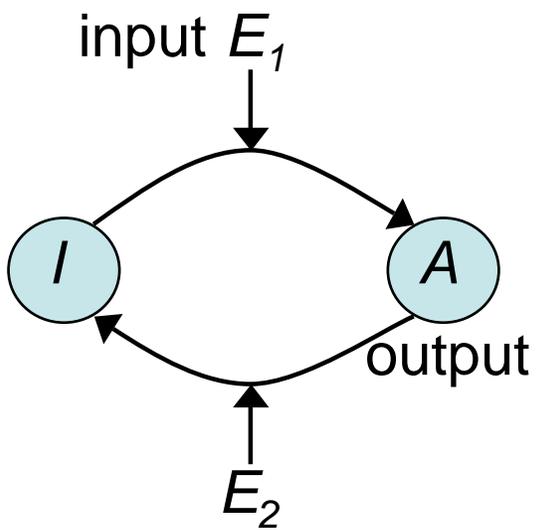

Forward reaction:

$$I + E_1 \underset{d_1}{\overset{a_1}{\rightleftarrows}} IE_1 \overset{k_1}{\longrightarrow} A + E_1$$

Backward reaction:

$$A + E_2 \underset{d_2}{\overset{a_2}{\rightleftarrows}} AE_2 \overset{k_2}{\longrightarrow} I + E_2$$

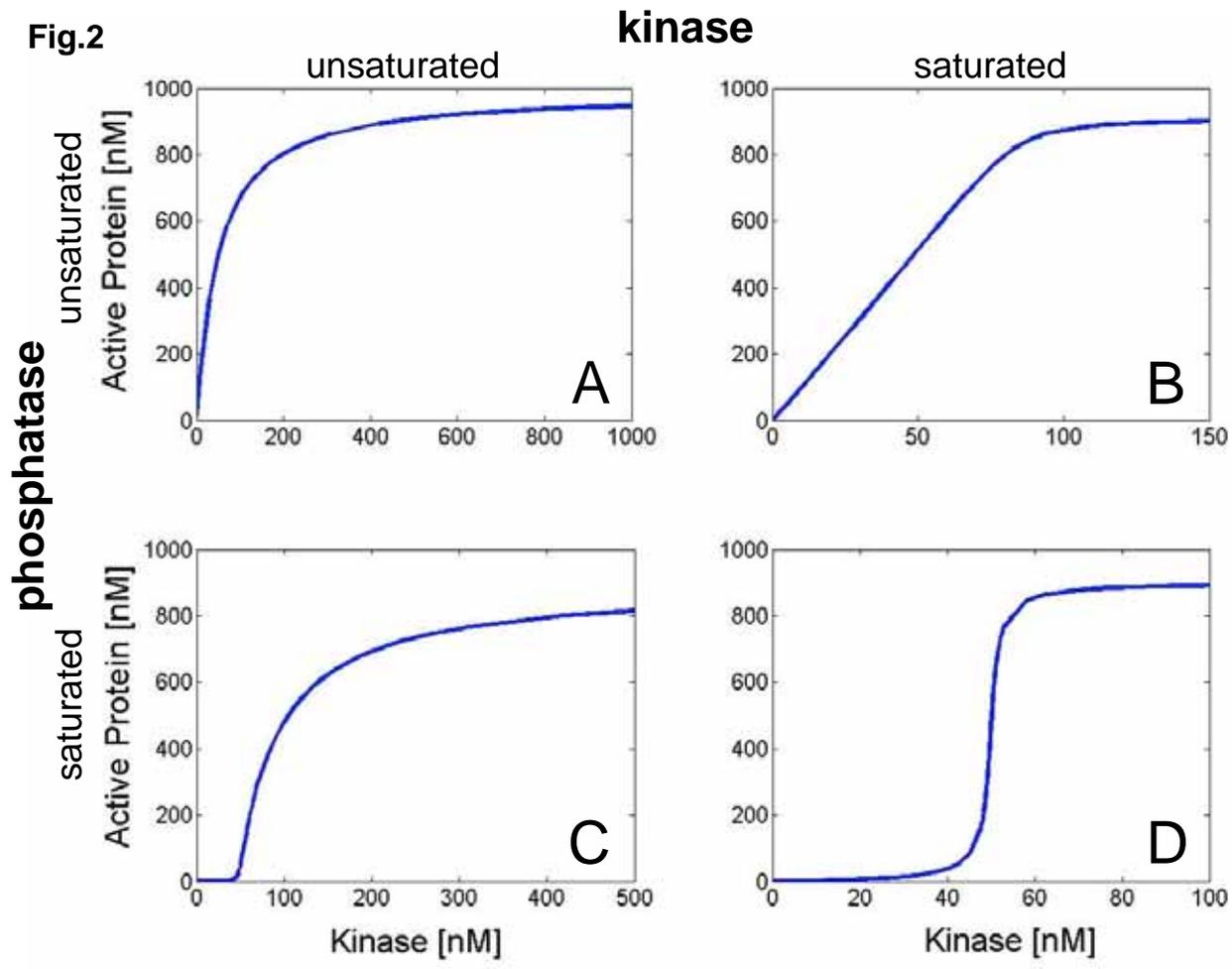

**Fig.3**

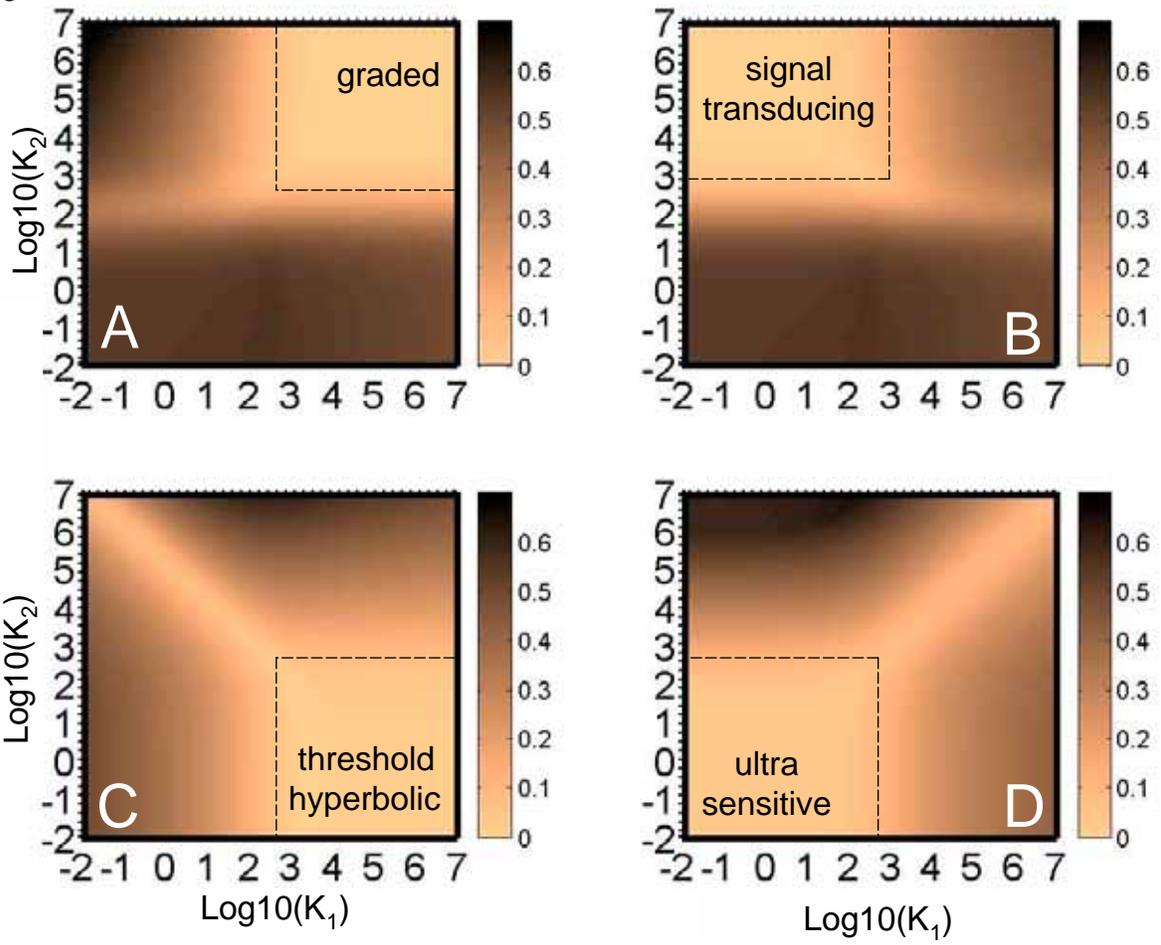

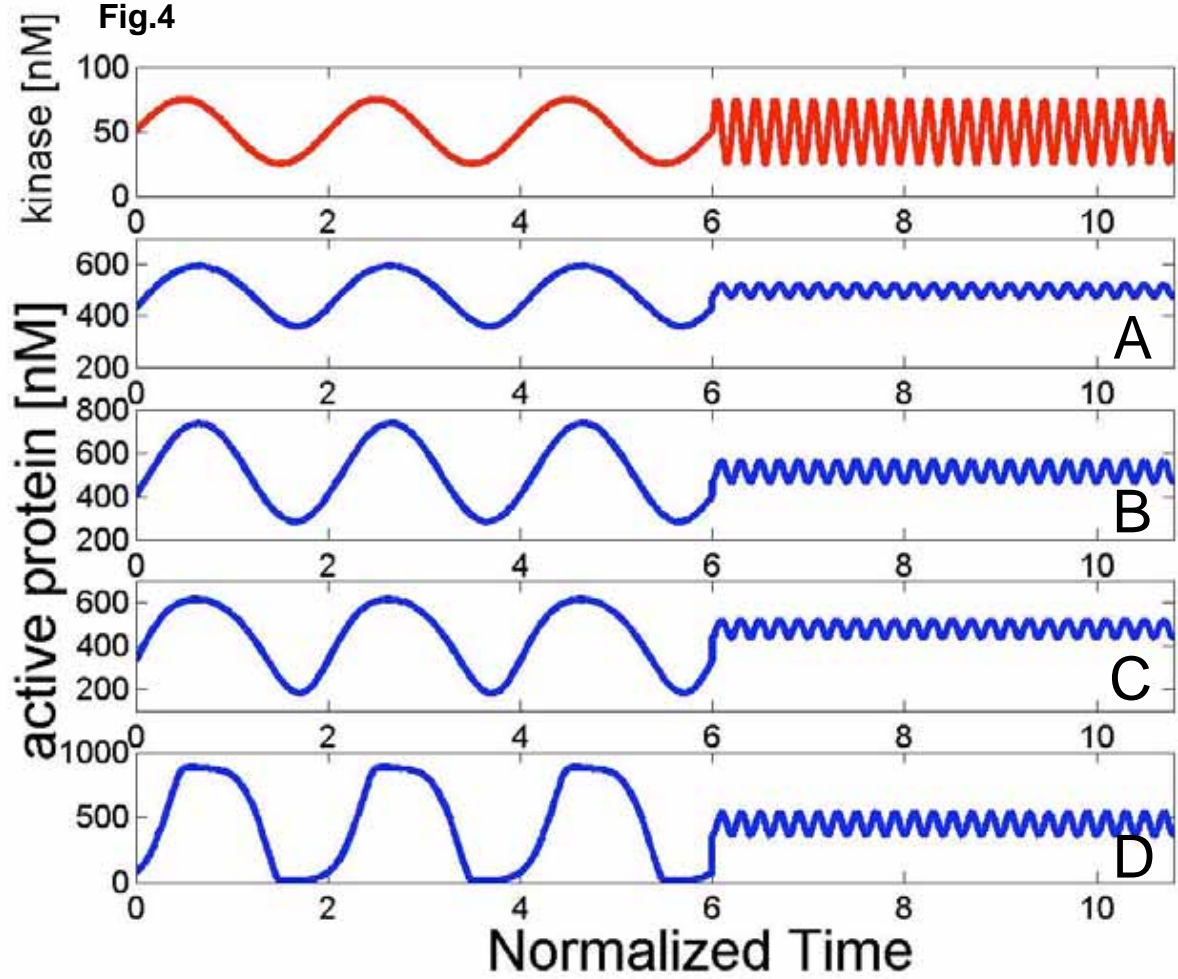

Fig.4

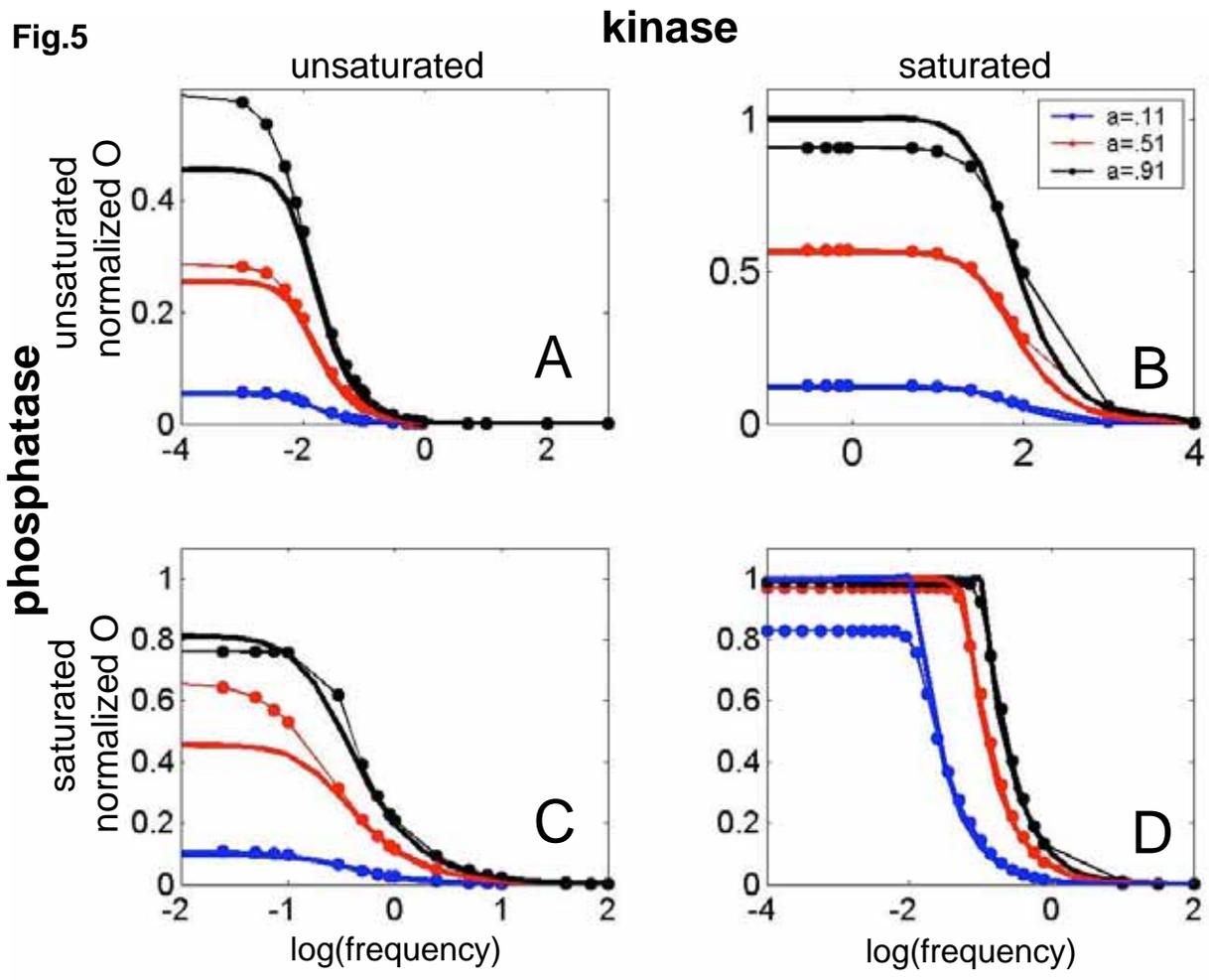

**Fig.6**

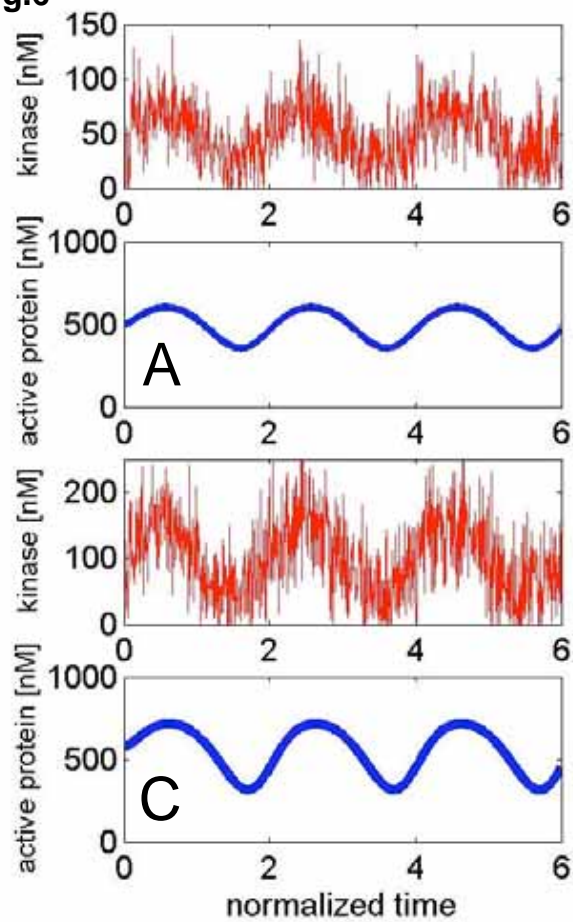
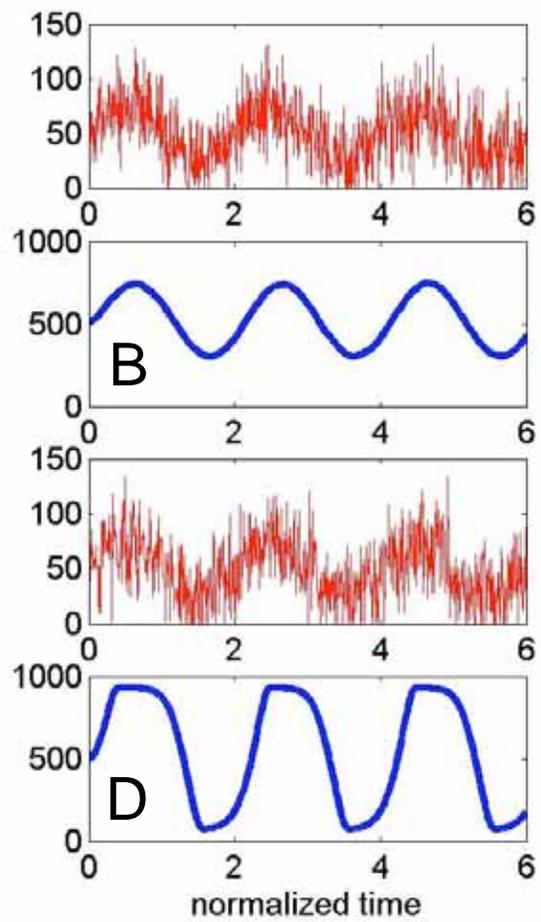